\documentclass[prd,amsfonts,noshowpacs,nofootinbib]{revtex4}
\usepackage{amsmath,graphicx}

\begin{document}
\title{Primordial non-Gaussianity from the DBI Galileons}
\author{Shuntaro Mizuno\footnote{shuntaro.mizuno@port.ac.uk}}
\author{Kazuya Koyama\footnote{Kazuya.Koyama@port.ac.uk}}
\affiliation{Institute of Cosmology and Gravitation, University of Portsmouth, Portsmouth PO1 3FX, UK.
}

\date{\today}
\begin{abstract}
We study primordial fluctuations generated during inflation
in a class of models motivated by the DBI Galileons, which 
are extensions of the DBI action that yield second order field equations. 
This class of models generalises the DBI Galileons in a similar way
with K-inflation. We calculate the primordial non-Gaussianity from the bispectrum of 
the curvature perturbations at leading order in the slow-varying approximations.
We show that the estimator for the equilateral-type non-Gaussianity, 
$f_{\rm NL} ^{equil}$, can be applied
to measure the amplitude of the primordial bispectrum
even in the presence of the
Galileon-like term although it gives a slightly
different momentum dependence from K-inflation models.
For the DBI Galileons, we find $-0.32 /c_s^2 < f_{\rm NL} ^{equil} < -0.16/c_s^2$
and large primordial non-Gaussianities can be
obtained when $c_s$ is much smaller than $1$ as in the usual DBI inflation.
In G-inflation models, where a de Sitter solution is obtained without any potentials, 
the non-linear parameter is given by $f_{\rm NL}^{equil} = 4.62 r^{-2/3}$ where $r$ is the 
tensor to scalar ratio, giving a stringent constraint on the model. 
 \end{abstract}

\maketitle

\section{Introduction}
The DBI inflation model \cite{Silverstein:2003hf}
is one of the most interesting possibilities to realise large
non-Gaussianity of the Cosmic Microwave Background (CMB)
temperature fluctuations.
Non-Gaussianity of the curvature perturbation in DBI
inflation has been studied extensively
\cite{Alishahiha:2004eh, Chen:2004gc,Chen:2005ad,Chen:2005fe,
Shandera:2006ax, Chen:2006nt,  Langlois:2008wt,
Langlois:2008qf, Arroja:2008yy, Khoury:2008wj,
Langlois:2009ej, Cai:2009hw,Huang:2006eh, Arroja:2008ga,
Arroja:2009pd, Chen:2009bc,Gao:2009gd, Mizuno:2009cv,
Mizuno:2009mv, RenauxPetel:2009sj, Gao:2009at}
(see also \cite{Koyama:2010xj, Chen:2010xk} for reviews).

Recently, a very interesting extension of
the DBI inflation model, so-called ``DBI Galileons'' was
proposed by de Rham and Tolley \cite{deRham:2010eu}
(see also \cite{Goon:2010xh}).
This is based on the relativistic extension of the
Galileon model \cite{Nicolis:2008in, Deffayet:2009wt,
Deffayet:2009mn} 
(for studies of cosmology based on the Galileon field, see
Refs.~\cite{Chow:2009fm,Silva:2009km,Kobayashi:2009wr,Kobayashi:2010wa,Gannouji:2010au,DeFelice:2010jn,
DeFelice:2010gb,Creminelli:2010ba,DeFelice:2010pv,Ali:2010gr,DeFelice:2010nf}).
The simplest example is a single field model that arises from a probe brane action in the five-dimensional spacetime. Let's consider the following four-dimensional induced action on the probe brane
\begin{equation}
S = \int d^4 x \sqrt{-g} \left(\lambda - M_5^3 K  \right),
\end{equation}
where $\lambda$ is a tension of the brane, $M_5$ is the five dimensional Planck constant, $g_{\mu \nu}$ is the induced metric on the brane,
\begin{equation}
g_{\mu \nu} = \eta_{\mu \nu} + \partial_{\nu} \pi \partial_{\mu} \pi,
\end{equation}
and $K$ is a trace of the extrinsic curvature $K_{\mu \nu}$,
\begin{equation}
K_{\mu \nu} = -\frac{\partial_{\mu} \partial_{\nu} \pi}{\sqrt{1 + (\partial \pi)^2}}.
\end{equation}
Here $\pi$ is a modulus describing the position of the brane.
Using the fact that the inverse metric is given by $g^{\mu \nu} =\eta^{\mu \nu} - \gamma^2
\partial^{\mu} \pi \partial^{\nu} \pi$ where $\gamma$ is the Lorentz factor
$\gamma = 1/\sqrt{1+ (\partial \pi)^2}$, the brane action is written as
\begin{equation}
S =- \lambda \int d^4 x \sqrt{1 + (\partial \pi)^2}
+ M_5^3  \int d^4 x \left( \Box \pi - \gamma^2 \partial_{\mu} \partial_{\nu} \pi
\partial^{\mu} \pi \partial^{\nu} \pi \right).
\end{equation}
By integrating by part and discarding the total derivative terms, this action can be rewritten as
\begin{equation}
S = - \lambda \int d^4 x \sqrt{1 + (\partial \pi)^2}
+ \frac{M_5^3}{2}  \int d^4 x \left(
\gamma^2 (\partial \pi)^2 \Box \pi + \partial_{\mu} (\gamma^2)
(\partial^{\mu} \pi) (\partial \pi)^2 \right).
\label{relgal}
\end{equation}
The first term is the usual DBI action. The higher order terms look containing the higher derivatives
but the equation of motion is at most second order in derivatives. In the non-relativistic limit
$\gamma \to 1$, this reduces the Galileon model where the action is invariant under the Galileon symmetry $\partial_{\mu} \pi \to \partial_{\mu} \pi + c_{\mu}$. It is also possible to include two more higher order terms in the action but in this paper we focus our attention to the leading order cubic order terms.

Recently, Refs.~\cite{Deffayet:2010qz} and
\cite{Kobayashi:2010cm} considered a generalisation of this model. This generalisation
is based on the extension of DBI inflation to K-inflation.
The generalised action is given by
\begin{equation}
S = \int d^4 x \sqrt{-g} (P(X, \phi) - G(X, \phi) \Box \phi)\,,
\label{Ginf}
\end{equation}
where $\phi$ is the same degree of the freedom as $\pi$,
but it is defined so that $\phi$ has a dimension of mass
and $X = -(1/2) (\partial \phi)^2$.
In Eq.~(\ref{Ginf}),
$P(\phi, X)$ and $G(\phi, X)$ are arbitrary functions
of $\phi$ and $X$.
Precisely speaking this generalisation does not include the action (\ref{relgal}). However, in the DBI inflation, the Lorentz factor, $\gamma$, varies very slowly and at leading order in the slow-varying parameters, which is usually used to calculate the leading order contribution to non-Gaussianity, the last term in (\ref{relgal}) does not play a role.  Thus the action (\ref{Ginf}) is general enough to include the case of the DBI Galileons. Again the equation of motion is at most second order in derivatives.

In this paper, we study primordial fluctuations generated during
inflation described by the action (\ref{Ginf}). We calculate the power
spectrum of the curvature perturbation as well as the bispectrum of the
curvature perturbations. Two examples of the model are considered: one
is the DBI Galileons described by the action (\ref{relgal}). The other is G-inflation models proposed by
Ref.~\cite{Kobayashi:2010cm}. This model is based on a specific choice of the functions $P(X)$ and $G(X)$ that realises a de Sitter solution without any potentials.

This paper is organised as follows. In section II, we introduce a model studied in
this paper. The power spectrum is calculated in section II
and the bispectrum is discussed in section III. In section IV, we
consider two examples, the DBI Galileons and G-inflation. Section V is devoted to conclusions.

\section{Model}

Assuming that $\phi$ described by the action (\ref{Ginf})
is minimally coupled to gravity,
we consider a class of models described
by the following action:
\begin{eqnarray}
&&S=\frac12 \int d^4 x \sqrt{-g} [M_{pl}^2 R + 2P(\phi, X)
-2 G(\phi,X) \Box \phi]\,,
\label{action}
\end{eqnarray}
where
$M_{pl}$ is the Planck mass.
In the background, we are interested in flat, homogeneous and
isotropic Friedmann-Robertson-Walker universes described by the
line element
\begin{equation}
ds^2=-dt^2+a^2(t)\delta_{ij}dx^idx^j, \label{FRW}
\end{equation}
where $a(t)$ is the scale factor.
It can be shown that the energy density and pressure of the field
are given by
\begin{eqnarray}
\rho &=& 2 P_{,X} X - P + 6 G_{,X} H \dot{\phi} X - 2 G_{,\phi} X\,,\\
p &=& P - 2(G_{,\phi} + G_{,X} \ddot{\phi}) X\,,
\end{eqnarray}
where $H = \dot{a}/a$ is the Hubble parameter,
a dot represents a derivative with respect to cosmic time
$t$ and the subscripts ${}_{,X}$ and ${}_{,\phi}$
denote derivatives with respect to
$X$  and $\phi$, respectively.
The Friedmann equation and the field equation are given by
\begin{eqnarray}
&&3 M_{pl}^2 H^2 = \rho\,,\label{bg_Friedmann}\\
&&P_{,X} (\ddot{\phi} + 3 H \dot{\phi}) + 2 P_{,XX} X \ddot{\phi}
+ 2 P_{,X \phi} X - P_{,\phi}
-2 G_{,\phi} (\ddot{\phi} + 3 H \dot{\phi})
-2 G_{, X \phi} X (\ddot{\phi} - 3 H \dot{\phi})\nonumber\\
&&+ 6 G_{,X} [(H X)^{\cdot} + 3 H^2 X] -2 G_{,\phi \phi}
+6 G_{,XX} H X \dot{X}=0\,.\label{bg_field_eq}
\end{eqnarray}

It is useful to define a slow-varying parameter
\begin{eqnarray}
\epsilon \equiv -\frac{\dot{H}}{H^2}=
\frac{X P_{,X} + 3 G_{,X} H \dot{\phi} X}
{M_{pl}^2 H^2}\,,
\label{def_epsilon}
\end{eqnarray}
where for the second equality, we have assumed the
quantities $|\ddot{\phi}/(H \dot{\phi})|$ and
$|G_{,\phi} \dot{\phi}/(G H)|$
are much smaller than $1$.
Since we are interested in fluctuations generated during
inflation, we will consider the background that 
satisfies the slow-varying conditions which are given by
$|\epsilon| \ll 1$, together with $|\ddot{\phi}/(H \dot{\phi})| \ll 1$ and
$|G_{,\phi} \dot{\phi}/(G H)| \ll 1$.

\section{Power spectrum}

We are interested in the primordial curvature perturbation on uniform
density hypersurfaces, $\zeta$, on large scales,
which is directly related to
temperature anisotropies in the Cosmic Microwave Background (CMB).
In order to calculate the statistical quantities
of $\zeta$ at leading order
in the slow-varying approximations, we first calculate the bispectrum of 
the fluctuations of inflaton $\phi$ in the flat gauge where the three-dimensional metric
takes the form $h_{ij} = a^2 \delta_{ij}$,
and then relate it to that of $\zeta$
using the relation obtained from the delta-N formalism
\cite{Maldacena:2002vr,Gruzinov:2004jx}
\begin{eqnarray}
\zeta=-\frac{H}{\dot{\phi}} Q\,.
\label{linear_rel_phi_zeta}
\end{eqnarray}
In Eq.~(\ref{linear_rel_phi_zeta}), $\phi$
is the background value and $Q$ is the perturbation
in the flat gauge.
In this gauge,
at leading order in the slow-varying approximations,
the second-order action is expressed as
\begin{eqnarray}
S_2 =\int dt d^3 x a^3 \biggl[P_{,X} \delta X^{(2)}+
 \frac12 P_{,XX} \left(\delta X^{(1)}\right)^2
 -G_{,X} \delta X^{(2)} \Box \phi^{(0)}
-G_{,X} \delta X^{(1)} \Box \phi^{(1)}
-\frac12 G_{,XX} \left(\delta X^{(1)}\right)^2\Box \phi^{(0)}
\biggr]\,,
\end{eqnarray}
with
\begin{eqnarray}
&&\delta X^{(1)} = \dot{\phi} \dot{Q}\,,
\hspace{0.5cm}
\delta X^{(2)} =\frac12 \dot{Q}^2 -\frac{1}{2 a^2}
\partial^i Q \partial_i Q\,,
\hspace{0.5cm}\Box \phi^{(0)} = -3H \dot{\phi}\,,
\hspace{0.5cm}\Box \phi^{(1)} = -\ddot{Q} - 3H\dot{Q}
+\frac{1}{a^2} \partial^i \partial_i Q\,.
\end{eqnarray}
Introducing the sound speed for the scalar perturbations
\begin{eqnarray}
c_s^2 = \frac{P_{,X} + 4 \dot{\phi} H G_{,X}}
{P_{,X} + 2 X P_{,XX} + 6 H \dot{\phi}
(G_{,X} + X G_{,XX})}\,,
\label{sound_speed}
\end{eqnarray}
and integrating by parts,
we can write the second order action as
\begin{eqnarray}
S_2 = \int dt d^3 x \frac{a^3}{2 c_s^2}
(P_{,X} + 4 \dot{\phi} H G_{,X})
\left[\dot{Q}^2
-\frac{c_s^2}{a^2} \partial^i Q \partial_i Q\right]\,.
\label{2nd_action}
\end{eqnarray}

The perturbations in the interaction picture are promoted to the quantum operators as
\begin{equation}
Q(\tau,\mathbf{x})=\frac{1}{(2\pi)^3}\int
d^3\mathbf{k}Q(\tau,\mathbf{k})e^{i\mathbf{k}\cdot\mathbf{x}}
\,,
\hspace{1cm}
Q(\tau,\mathbf{k})=u(\tau,\mathbf{k})a(\mathbf{k})+u^*(\tau,-\mathbf{k})a^\dag(-\mathbf{k})\,,
\end{equation}
where $a(\mathbf{k})$ and $a^\dag(-\mathbf{k})$ are the annihilation
and creation operator respectively. They satisfy the usual
commutation relations
\begin{equation}
\left[a(\mathbf{k_1}),a^\dag(\mathbf{k_2})\right]
=(2\pi)^3\delta^{(3)}(\mathbf{k_1}-\mathbf{k_2})\,,
\hspace{1cm}
\left[a(\mathbf{k_1}),a(\mathbf{k_2})\right]=\left[a^\dag(\mathbf{k_1}),a^\dag(\mathbf{k_2})\right]=0\,.
\end{equation}
From the second order action (\ref{2nd_action}), the solution for the mode functions is given by
\begin{equation}
u(\tau,\mathbf{k})=
\frac{H}{\sqrt{2 c_s (P_{,X} + 4 \dot{\phi} H G_{,X})}}
\frac{1}{k^{3/2}}\left(1+ikc_s\tau\right)e^{-ikc_s\tau}\,.
\end{equation}
It is convenient to introduce the following parameters
\begin{equation}
\nu \equiv \frac{G_{,X} \dot{\phi} X}{M_{pl}^2 H}\,,
\hspace{0.5cm}
\tilde{\epsilon} \equiv \epsilon +\nu\,,
\label{def_tilde_epsilon}
\end{equation}
where $\tilde{\epsilon}$
coincides with $\epsilon$
when there is no Galileon-like term.
$\tilde{\epsilon}$ is also much smaller than $1$ for
$\nu \ll 1$.
Then,
the power spectrum of $Q$ and $\zeta$
are given by
\begin{eqnarray}
&&\langle Q (\mathbf{k_1}) Q (\mathbf{k_2}) \rangle
= (2 \pi)^3 \delta^{(3)} (\mathbf{k_1} + \mathbf{k_2})
\mathcal{P}_Q \frac{2 \pi^2}{k_1^3}\,,\;\;\;\;\;
\mathcal{P}_Q = \frac{X}{4 \pi^2 M_{pl}^2 c_s
\tilde{\epsilon}}
\label{power_spectrum_q}\,,\\
&&\langle \zeta (\mathbf{k_1}) \zeta (\mathbf{k_2}) \rangle
= (2 \pi)^3 \delta^{(3)} (\mathbf{k_1} + \mathbf{k_2})
\mathcal{P}_\zeta \frac{2 \pi^2}{k_1^3}\,,\;\;\;\;\;
\mathcal{P}_\zeta =
\frac{1}{8 \pi^2 M_{pl}^2}
\frac{H^2}{c_s \tilde{\epsilon}}
\label{power_spectrum_zeta}\,,
\end{eqnarray}
which are evaluated
at the time of the sound horizon exit, $c_s k = a H$.
It is worth noting that it is $\tilde{\epsilon}$
not $\epsilon$ which appears in
Eq.~(\ref{power_spectrum_zeta}), which gives the
behaviour of the power spectrum that is different from
the usual K-inflation model as we will see below.
Defining additional slow-varying parameters
\begin{eqnarray}
\tilde{\eta} \equiv \frac{\dot{\tilde{\epsilon}}}
{\tilde{\epsilon} H}\,,\hspace{0.5cm}
s \equiv \frac{\dot{c_s}}{c_s H}\,,
\end{eqnarray}
the spectral index of the primordial power spectrum
is given by
\begin{eqnarray}
n_s-1 = \frac{d \ln \mathcal{P}_\zeta (k)}{d \ln k}
= -2 \epsilon - \tilde{\eta} - s\,.
\label{spectral_index}
\end{eqnarray}
We need to require $\epsilon$,
$\tilde{\eta}$, $s$ to be very small
in order to realise the almost scale invariant power spectrum.
We have confirmed that this result is consistent with
the one obtained in Ref.~\cite{Kobayashi:2010cm}
when the conditions
$|\ddot{\phi}/(H \dot{\phi})| \ll 1$ and
$|G_{,\phi} \dot{\phi}/(G H)| \ll 1$ are satisfied.
Notice that $\tilde{\eta}$ is different from the usual $\eta$
defined by $\eta \equiv \dot{\epsilon}/(\epsilon H)$.

The power spectrum and spectral index of tensor perturbations
are given by the usual expression
\begin{eqnarray}
\mathcal{P}_T =
\frac{2 H^2}{\pi^2 M_{pl}^2}\,,\hspace{0.5cm}
 n_T = -2 \epsilon\,.
\label{power_spectrum_h}
\end{eqnarray}

In the usual K-inflation, $n_T$ and tensor to scalar ratio
$r \equiv \mathcal{P}_T/\mathcal{P}_\zeta$ are
not independent, and there is a so-called
``consistency relation'' $r=-8c_s n_T$ \cite{Garriga:1999vw}.
However, it is clear that this relation does not hold
in the presence of the Galileon-like term.
Instead, we have
\begin{eqnarray}
r = - 8c_s (n_T - 2 \nu)\,,
\label{tensor_to_scalar}
\end{eqnarray}
where $\nu$ is given by Eq.~(\ref{def_tilde_epsilon}).

\section{Bispectrum}

The third order action can be obtained in the same way as 
\begin{eqnarray}
S_3 &=&\int dt d^3 x \frac{a^3}{\dot{\phi}}\biggl[C_1
 \dot{Q}^3+\frac{C_2}{a^2} \dot{Q} \partial^i Q \partial_i Q
+\frac{C_3}{a^4 H} \partial^i Q \partial_i Q
\partial^j \partial_j Q+\frac{C_4}{a^2H} \dot{Q}
\partial^i \dot{Q} \partial_i Q\biggr]\,,
\label{leadingorderaction}
\end{eqnarray}
where
\begin{eqnarray}
&&C_1=\frac23 X^2 P_{,XXX}+
X P_{,XX} + 2 H \dot{\phi} X^2 G_{,XXX} +
5H \dot{\phi} X G_{,XX}
+H \dot{\phi} G_{,X}\,,\nonumber\\
&&C_2=-\left(X P_{,XX} +
3H \dot{\phi} X G_{,XX} + H \dot{\phi} G_{,X}\right)\,,
\hspace{0.5cm}
C_3=\frac12 H \dot{\phi} G_{,X}\,,
\hspace{0.5cm}C_4=2H\dot{\phi} G_{,X}
+ 2H\dot{\phi} X G_{,XX}\,.
\label{abcde}
\end{eqnarray}

The vacuum expectation value of the three point operator in the
interaction picture  is written as
\cite{Maldacena:2002vr,Weinberg:2005vy}
\begin{equation}
\langle Q(t,\mathbf{k_1})Q(t,\mathbf{k_2})
Q(t,\mathbf{k_3})\rangle=-i\int^t_{t_0}d\tilde
t \langle
\left[Q(t,\mathbf{k_1})Q(t,\mathbf{k_2})Q(t,\mathbf{k_3}),
H_I(\tilde
t)\right] \rangle\,,
\end{equation}
where $t_0$ is some early time during inflation
when the field's vacuum fluctuation
are deep inside the sound horizon and 
$t$ is some time after the sound horizon exit.
If one uses a conformal time, it is a good approximation
to perform the integration from $-\infty$ to $0$
because $\tau\approx-(aH)^{-1}$.
$H_I$ denotes the interaction Hamiltonian and it is given by
$H_I=-L_3$, where $L_3$ is the lagrangian obtained
from the action (\ref{leadingorderaction}).
Using the solution for the mode function and the 
commutation relations for the creation and annihilation
operators, we get
\begin{eqnarray}
\langle Q(\mathbf{k_1})Q(\mathbf{k_2})
Q(\mathbf{k_3})\rangle = (2 \pi)^3
\delta^{(3)}(\mathbf{k_1}+\mathbf{k_2}+\mathbf{k_3} )
\frac{H^5}{(P_{,X} + 4 \dot{\phi} H G_{,X})^3 \dot{\phi}}
\frac{1}{\Pi _{i=1} ^3 k_i ^3}\mathcal{A}_\phi\,,
\end{eqnarray}
where
\begin{eqnarray}
\mathcal{A}_\phi&=&3
\left(C_1 - \frac{ C_4}{c_s^2}\right) \mathcal{A}_1
+ \frac{ C_2}{2 c_s^2} \mathcal{A}_2 +
\frac{ C_3}{c_s^4}\mathcal{A}_3\,,
\label{quant_bispectrum}
\end{eqnarray}
and $C_1$, $C_2$, $C_3$, $C_4$ and $C_5$
are given by Eq.~(\ref{abcde}).
In Eq.~(\ref{quant_bispectrum}), we have introduced
the shape functions $\mathcal{A}_1$, $\mathcal{A}_2$ and 
$\mathcal{A}_3$ as
\begin{eqnarray}
\mathcal{A}_1&=&\frac{k_1 ^2 k_2 ^2 k_3 ^2}{K^3}\,,
\label{shape_func_a1}\\
\mathcal{A}_2&=&
\frac{k_1^2 \mathbf{k_2} \cdot \mathbf{k_3}}{K}
\left(1+\frac{k_2 + k_3}{K}+2 \frac{k_2 k_3}{K^2}\right)
+2\;{\rm perms.}\,,
\label{shape_func_a2}\\
\mathcal{A}_3&=&
\frac{k_3^2 \mathbf{k_1} \cdot \mathbf{k_2} }{K}
\left(1+\frac{k_1 k_2+k_2 k_3 + k_3 k_1}{K^2}
+3\frac{k_1 k_2 k_3}{K^3}\right)
+2\;{\rm perms.}\,,
\label{shape_func_a3}
\end{eqnarray}
where $K=k_1+k_2+k_3$.
The shapes $\mathcal{A}_1$ and $\mathcal{A}_2$ appear
in the usual K-inflation models \cite{Chen:2006nt}
and their amplitudes can be measured by the estimator 
for the equilateral-type non-Gaussianity,
$f_{\rm NL} ^{equil}$ \cite{Babich:2004gb}.
On the other hand,
the shape $\mathcal{A}_3$ is completely new that arises from the Galileon-like term.

For the bispectrum of $\zeta$,
making use of Eqs.~(\ref{linear_rel_phi_zeta})
and (\ref{power_spectrum_zeta}),
we obtain the following expression:
\begin{eqnarray}
&&\langle \zeta(\mathbf{k_1}) \zeta(\mathbf{k_2})
\zeta(\mathbf{k_3}) \rangle
= (2 \pi)^3
\delta^{(3)}(\mathbf{k_1}+\mathbf{k_2}+\mathbf{k_3} )
(\mathcal{P}_{\zeta})^2  F(k_1, k_2, k_3) \,,
\label{bispectrum_gen}
\end{eqnarray}
where
\begin{eqnarray}
&&F(k_1, k_2, k_3) =\frac{(2 \pi)^4}{\Pi _{i=1} ^3 k_i ^3}
\sum_{j=1} ^3 f ^{(j)} \mathcal{A}_j\,,
\label{shape_gen}\\
&&f ^{(1)} =
\frac{-3 (C_1 c_s^2-C_4)}{(P_{,X} + 4\dot{\phi} H G_{,X})}\,,
\;\;\;\;
f ^{(2)} =
-\frac{C_2}{2(P_{,X} + 4\dot{\phi} H G_{,X})}\,,
\;\;\;\;f ^{(3)} =
-\frac{C_3}{(P_{,X} + 4\dot{\phi} H G_{,X})c_s^2}\,.
\label{gen_f}
\end{eqnarray}

In the following, we will
study the momentum dependence of the bispectrum.
Especially, we will
check the validity of using the estimator for equilateral type non-Gaussianity,
$f_{NL} ^{equil}$, for the bispectrum even in the presence of the new shape $\mathcal{A}_3$. 
The estimator is defined by
\begin{eqnarray}
F^{equil}(k_1, k_2, k_3) = (2 \pi)^4 \left(\frac{9}{10} f_{\rm NL} ^{equil}\right)
\left(-\frac{1}{k_1^3 k_2^3}-\frac{1}{k_1^3 k_3^3} -\frac{1}{k_2^3 k_3^3}
-\frac{2}{k_1^2 k_2^2 k_3^2} + \frac{1}{k_1 k_2^2 k_3^3 } + (5\;perms.)\right)\,,
\label{equil_shape}
\end{eqnarray}
where the permutations act only on the last term in parentheses. This shape is 
factorisable and it is possible to construct a fast optimal estimator that can be 
applied to the CMB map. For this purpose, it is useful to define
shape functions $F^{(i)} (k_1, k_2, k_3)$, $i=1,2,3$ corresponding to the
shapes $\mathcal{A}_i$ in Eq.~(\ref{shape_gen}).
As mentioned before, the shape
of the bispectrum in K-inflation is characterised by the sum of
$F^{(1)} (k_1, k_2, k_3)$ and $F^{(2)} (k_1, k_2, k_3)$
with $f^{(1)}$
and  $f^{(2)}$ depending on $P(X)$.
For example, in DBI inflation,
the relation $f^{(2)}/f^{(1)}=-2/3$ holds. However, since functions 
$F^{(1)} (k_1, k_2, k_3)$ and $F^{(2)}(k_1, k_2, k_3)$ are not factorisable,
$F^{equil} (k_1, k_2, k_3)$ is usually used to 
approximate the shape functions 
$F^{(1)} (k_1, k_2, k_3)$ and $F^{(2)} (k_1, k_2, k_3)$.

\begin{figure}[h]
\scalebox{0.6}{
\centerline{
\includegraphics{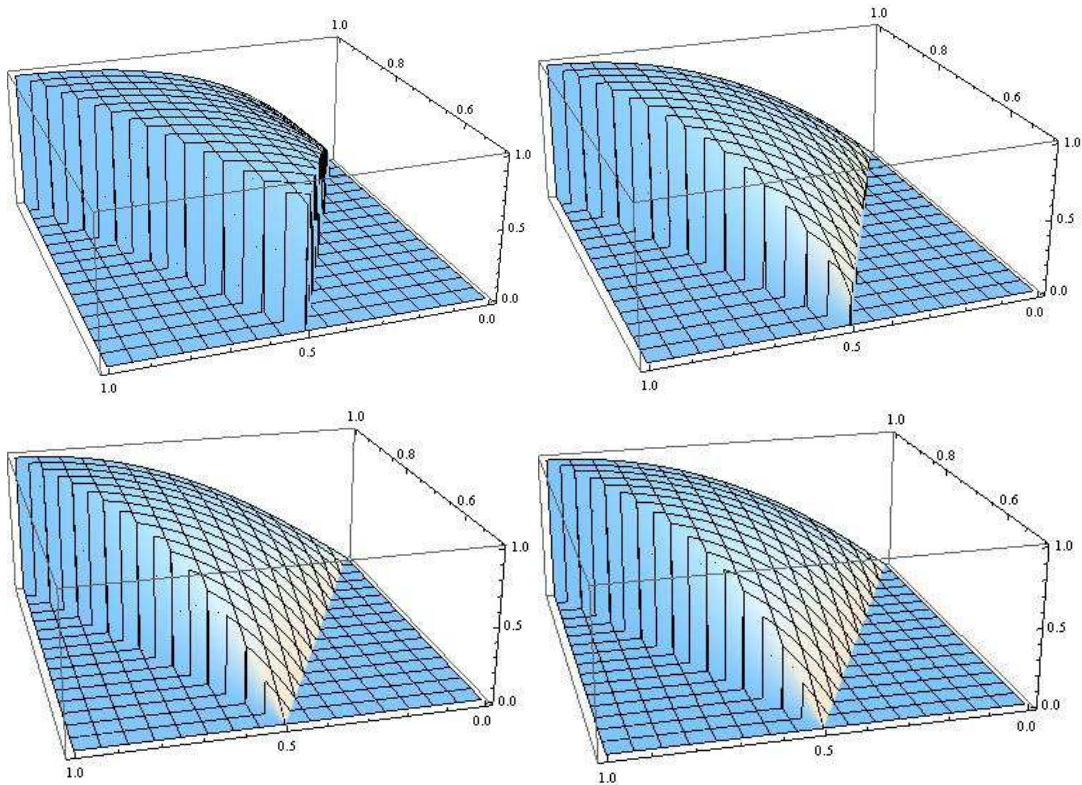}}}
\caption{
In this group of figures, we plot
the shape functions $F(1,k_2/k_1, k_3/k_1) (k_2/k_1)^2
(k_3/k_1)^2$ as functions of $(k_2/k_1, k_3/k_1)$. 
The figures are normalised to have values $1$
for equilateral configurations $k_2/k_1 = k_3/k_1 =1$
and set to zero outside the region $1-k_2/k_1 \leq k_3/k_1
\leq k_2/k_1$. We plot $F^{(1)} (k_1, k_2, k_3)$,
$F^{(2)} (k_1, k_2, k_3)$, $F^{(3)}(k_1, k_2, k_3)$
and $F^{equil} (k_1, k_2, k_3)$ for upper left, upper right,
lower left, lower right, respectively.}
\label{shapefunctions}
\end{figure}

In Figure~\ref{shapefunctions},
we compare $F^{equil}(k_1, k_2, k_3)$ with
$F^{(1)} (k_1, k_2, k_3)$, $F^{(2)} (k_1, k_2, k_3)$
and $F^{(3)} (k_1, k_2, k_3)$
with appropriate normalisations.
From this, we see that not only
$F^{(1)} (k_1, k_2, k_3)$ and $F^{(2)} (k_1, k_2, k_3)$,
but also the function $F^{(3)} (k_1, k_2, k_3)$
has a very similar shape with $F^{equil} (k_1, k_2, k_3)$
and it can be expected that $f^{equil} _{\rm NL}$
provides a good measure of the bispectrum even in the presence of 
the new shape function $F^{(3)}$.
We can prove this quantitatively by the following
shape correlator $\mathcal{C}$ for two different
shapes characterised by $F$ and $F'$
introduced by Ref.~\cite{Fergusson:2008ra}
(See also \cite{2010arXiv1004.2915R, Mizuno:2010by}),
\begin{eqnarray}
\mathcal{C} (F, F') = \frac{\mathcal{F} (F, F')}
{\sqrt{\mathcal{F} (F, F) \mathcal{F} (F', F')}}\,,
\label{shape_correlator_b}
\end{eqnarray}
where the overlap function $\mathcal{F}$ is given by
\begin{eqnarray}
\mathcal{F} (F, F') = \int d \mathcal{V}_k
F(k_1, k_2, k_3) F'(k_1, k_2, k_3) \Pi_{i=1} ^{4} k_i ^4
w_B(k_1, k_2, k_3)\,.
\label{pribispectrum_correlator_part}
\end{eqnarray}
In Eq.~(\ref{pribispectrum_correlator_part})
the integration is performed for the region
where the triangle condition for $(k_1, k_2, k_3)$ holds
and weight function $w_B$ is given by
\begin{eqnarray}
w_B= \frac{1}{k_1 + k_2 + k_3}\,.
\end{eqnarray}

Table~\ref{table_shape_correlation} shows that
the shapes $F^{equil} (k_1, k_2, k_3)$
and $F^{(3)} (k_1, k_2, k_3)$
are almost completely anti-correlated, which means
that it is very reasonable to adopt the estimator
$f_{\rm NL} ^{equil}$ to measure the amplitude of
the bispectrum even in the presence of
the shape $F^{(3)} (k_1, k_2, k_3)$.

\begin{widetext}
\begin{center}
\begin{table} [h!]
\begin{tabular} { |c|c|c|c|c| }
\hline
 & $F^{(1)} (k_1, k_2, k_3)$ & $F^{(2)} (k_1, k_2, k_3)$
& $F^{(3)} (k_1, k_2, k_3)$ &$F^{DBI} (k_1, k_2, k_3)$ \\
\hline
Overlap& $0.936$ & $-0.995$  & $-0.99989$ & $-0.993$\\
\hline
\end{tabular}
\caption{Shape correlations between the factorisable
equilateral shape $F^{equil} (k_1, k_2, k_3)$
and the shapes of primordial bispectra characterised by
functions $F^{(1)} (k_1, k_2, k_3)$, $F^{(2)} (k_1, k_2, k_3)$
and $F^{(3)} (k_1, k_2, k_3)$.
For comparison, we have also shown the correlation between
the equilateral shape and that obtained in DBI inflation
which is given by
$F^{(1)} (k_1, k_2, k_3)+F^{(2)} (k_1, k_2, k_3)$ satisfying
$f^{(2)}/f^{(1)}=-2/3$.}
\label{table_shape_correlation}
\end{table}
\end{center}
\end{widetext}

Now that the validity of using the estimator 
for the equilateral-type non-Gaussianity,
$f_{NL} ^{equil}$ for the bispectrum corresponding to
the new shape $\mathcal{A}_3$ is confirmed, we will
obtain $f_{\rm NL} ^{equil}$ for the bispectrum
given by Eq.~(\ref{shape_gen}).
We match the two different shapes (\ref{shape_gen}) and
(\ref{equil_shape}) so that these two shapes have the same
amplitudes at the equilateral configuration $k_1=k_2=k_3$.
This gives 
\begin{eqnarray}
f_{\rm NL} ^{equil} = \frac{10}{243} f^{(1)} -
\frac{85}{81} f^{(2)} - \frac{65}{81}f^{(3)}\,.
\label{fnl_gen_simplematching}
\end{eqnarray}

\section{Examples}

\subsection{The DBI Galileons}

As a first example, we consider
the DBI Galileons \cite{deRham:2010eu}
described by the action (\ref{relgal}). 
Here we extended the original DBI Galileon model by
introducing $V(\phi)$, $f(\phi)$ and $g(\phi)$ 
which depend on $\phi$ weakly so that the slow-varying
parameters are sufficiently small.
\begin{eqnarray}
P(\phi,X)=-f(\phi)^{-1} \sqrt{1-2Xf(\phi)} + f(\phi)^{-1}
-V(\phi)\,,\;\;
G(\phi,X)=\frac{g(\phi)X}{1-2Xf(\phi)}\,.
\label{p_g_dbi}
\end{eqnarray}

It is worth mentioning that 
only particular choices of the functions $f(\phi)$
and $g(\phi)$ come from genuine higher dimensional
symmetries, although
the equations of motions
are still kept to be second order.

In order to analyse this model, it is convenient to
define $c_D \equiv 1/ P_{,X}$ which corresponds to the sound speed
in the DBI model in the absence of the Galileon-like term.
Since it is known that
$f_{NL} ^{equil} \propto 1/c_D^2$ in DBI inflation and
we are interested in the case where
the large non-Gaussianity is generated,
we consider only the case with $c_D \ll 1$.

We assume that the inflation is
driven by the potential term, that is,
$V \gg X/c_D, g H \dot{\phi} X$
in Eq.~(\ref{bg_Friedmann}) as in usual DBI inflation. 
During inflation,
the field equation (\ref{bg_field_eq}) becomes
\begin{eqnarray}
\frac{3}{c_D} H\dot{\phi} + \frac{18 g H^2 X}{c_D^4}
+ V_{,\phi}=0\,.
\label{field_eq_dbi}
\end{eqnarray}
The value of $c_D$ is specified from the background equations once
$V(\phi)$ and $f(\phi)$ are given. We assume $\dot{c_D}/c_D H \ll 1$ 
so that, at the leading order, $c_D$ is constant (it is also possible 
to construct a model where $c_D$ is constant by choosing a functional
form of $f(\phi)$ for a given $V(\phi)$, see Ref.~\cite{Copeland:2010jt} 
for the details).
We define a parameter $b_D \equiv g  H \dot{\phi}/c_D^3$
so that the first two terms in Eq.~(\ref{field_eq_dbi})
are comparable when $b_D$ is of order $1$.
It is worth noting that $b_D$ can be also expressed as
$b_D = \nu/(\epsilon - 3 \nu)$.
Making use of the concrete forms of $P(X,\phi)$ and
$G(X, \phi)$ in Eq.~(\ref{p_g_dbi}) as well as
the definition of $b_D$,
the coefficients $C_1$, $C_2$, $C_3$ and $C_4$
in the third order action (\ref{abcde}) are given by
\begin{eqnarray}
C_1 = \frac{1}{2 c_D^5} (1+24 b_D)\,,
\hspace{0.5cm}
C_2=-\frac{1}{2 c_D^3}(1+12 b_D)\,,
\hspace{0.5cm}
C_3=-\frac{1}{2 c_D}b_D\,,
\hspace{0.5cm}
C_4=\frac{4}{c_D^3} b_D\,,
\end{eqnarray}
where we have used
$X P_{,XX} \sim 1/(2 c_D^3)$,
$X^2 P_{,XXX} \sim 3/(4 c_D^5)$, $G_{,X} \sim g/c_D^4$,
$XG_{,XX} \sim 2 g /c_D^6$, $X^2 G_{,XXX} \sim 6 g /c_D^8$
and $c_D \ll 1$.

Then,
the parameters $f^{(j)}$ in Eq.~(\ref{gen_f})
become
\begin{eqnarray}
f ^{(1)} = -\frac{3}{2 c_D^2}\frac{(1+20b_D)}{(1+4 b_D)(1+12 b_D)}\,,\;\;\;\;
f ^{(2)} = \frac{(1+12 b_D)}{4 (1+4b_D) c_D^2}\,,\;\;\;\;
f ^{(3)} = -\frac{(1+12 b_D)b_D}{2(1+4 b_D)^2 c_D^2}\,.
\label{f_dbi}
\end{eqnarray}
From Eq.~(\ref{fnl_gen_simplematching}) we obtain
\begin{eqnarray}
f_{\rm NL} ^{equil} =
-\frac{5}{324c_s^2}
\frac{(21 + 546 b_D +3776 b_D^2 + 6048b_D^3)}
{(1+4 b_D) (1+12 b_D)^2}\,,
\label{fnl_dbi}
\end{eqnarray}
which becomes $-0.16 /c_s^{2}$ for $b_D \gg 1$
and $-0.32/ c_s^{2}$ for $b_D \to 0$.
To obtain Eq.~(\ref{fnl_dbi}), we have used that $c_D$
and $c_s$ are related as
\begin{eqnarray}
c_D^2 = \frac{(1+12 b_D)}{(1+4 b_D)} c_s^2\,.
\end{eqnarray}
The nonlinear parameter $f_{\rm NL}^{equil}$ scales as
$\propto 1/c_s^2$ and can be detectable by future experiments such as PLANCK for sufficiently small $c_s$.

Especially, in the case of $\tilde{\eta}=s=0$,
we find that the following relation holds:
\begin{eqnarray}
r = 8 c_s \frac{1+4 b_D}{1+3 b_D} (1-n_s)\,.
\label{r_dbi}
\end{eqnarray}
Combined this with Eqs.~(\ref{r_dbi}) and (\ref{fnl_dbi}),
we can express $f_{\rm NL} ^{equil}$
in terms of $n_s$ and $r$ as
\begin{eqnarray}
f_{\rm NL} ^{equil} \simeq -20 \frac{(1-n_s)^2}{r^2}\,,
\end{eqnarray}
where the coefficient is almost independent of $b_D$
for this set up.
This relation suggests that these 
Galileon-like terms do not help 
embed the DBI inflation into string theory
\cite{Baumann:2006cd, Lidsey:2007gq}.

\subsection{G-inflation}

The second example is recently proposed G-inflation where
an exact de Sitter solution is realised
without introducing a potential \cite{Kobayashi:2010cm}.
In this model the functions $P(X)$ and $G(X)$ are chosen as 
\begin{eqnarray}
P(X)=-X + \frac{X^2}{2 M^3 \mu}\,,\;\;
G(X)=\frac{1}{M^3} X\,.
\label{p_g_ginf}
\end{eqnarray}
The de Sitter solution is obtained when $P_{,X}+3 H \dot{\phi}/M^3=0$
is satisfied in Eq.~(\ref{bg_Friedmann}). The solutions are obtained as 
\begin{eqnarray}
X = \mu M^3 x\,,\;\;\;\;H^2 = \frac{M^3}{18\mu} \frac{(1-x)^2}{x}\,,\;\;\;\;
{\rm where}\;\;\;\;\frac{1-x}{x \sqrt{1-x/2}} = \sqrt{6} \frac{\mu}{M_{pl}}\,.
\label{g-inf_def_x}
\end{eqnarray}
Here as in Ref.~\cite{Kobayashi:2010cm}, we only consider
simple cases where $\mu \ll M_{pl}$. 
In this situation, $(1-x) \simeq \sqrt{3} \mu /M_{pl} \ll 1$ and
$\mu$ is related with $M_{pl}$, $M$, $H$ as $\mu = 6 M_{pl}^2 H^2 /M^3$.

Then,  making use of the concrete forms of $P(X,\phi)$ and
$G(X, \phi)$ in Eq.~(\ref{p_g_ginf}), the fine-tuning condition
$P_{,X}+3 H \dot{\phi}/M^3=0$ and Eq.~(\ref{g-inf_def_x}),
the primordial power spectrum and tensor to scalar ratio are given by
\begin{eqnarray}
\mathcal{P}_{\zeta} = \frac{\sqrt{6} H^2}{16 \pi^2 M_{pl}^2}
\frac{1}{(1-x)^{3/2}}
\,,\hspace{0.5cm}r = \frac{16\sqrt{6}}{3} (1-x)^{3/2}\,,
\label{g_inf_power}
\end{eqnarray}
where we neglected the correction to the tensor to scalar ratio, $r=16 c_s \tilde{\epsilon}$,
that arises if we allow a small deviation from the pure de Sitter inflation. 
As was pointed by Ref.~\cite{Kobayashi:2010cm},
$r$ becomes nonzero even in the pure de Sitter solution, i.e. for $\epsilon=0$.
On the other hand, $n_s-1$ becomes $0$ for the pure de Sitter inflation, but
again if we allow a small deviation for it parameterised by the slow-varying parameters,
$n_s$ is given by Eq.~(\ref{spectral_index}).

For the primordial bispectrum,
the parameters $f ^{(j)}$
in Eq.~(\ref{gen_f})
become
\begin{eqnarray}
f ^{(1)} = \frac{9}{2}\,,\;\;\;\;
f ^{(2)} = \frac{3}{2(1-x)}\,,\;\;\;\;
f ^{(3)} = -\frac{3}{(1-x)}\,,
\label{f_ginf}
\end{eqnarray}
where we have set $x=1$ unless it appears
in the form of $1-x$. Then 
from Eq.~(\ref{fnl_gen_simplematching}) we obtain
\begin{eqnarray}
f_{\rm NL} ^{equil} = \frac{5}{6(1-x)}\,.
\label{fnl_ginf}
\end{eqnarray}
Especially, from Eq.~(\ref{g_inf_power}) and assuming the pure de Sitter inflation, $\epsilon=0$, 
$f_{\rm NL} ^{equil}$ is related with $r$ as
\begin{eqnarray}
f_{\rm NL} ^{equil}=
4.62  \frac{1}{r^{2/3}}\,,
\end{eqnarray}
which gives a strong constraint on this model.
For example if $r=0.17$, which can be detected
by the PLANCK satellite \cite{Planck},
$f_{\rm NL} ^{equil} = 15.1$. 
Therefore, a detection or non-detection of the tensor mode 
and equilateral type non-Gaussianity by PLANCK will tightly constrain the 
model.

\section{Conclusion}

The DBI inflation model \cite{Silverstein:2003hf}
has been extensively studied recently for both
theoretical and phenomenological reasons.
Especially, it is well known that the DBI inflation
model can give large non-Gaussianity
of the Cosmic Microwave Background (CMB) temperature
fluctuations. Recently, de Rham and Tolley \cite{deRham:2010eu} proposed
a new model so called the DBI Galileons
based on a probe brane action in the higher-dimensional
space time. Interestingly, this model
naturally provides a connection between the DBI model
and the relativistic generalisation of the Galileon model \cite{Nicolis:2008in} 
where the equation of motion is at most second order 
in derivatives due to the Galileon symmetry 
$\partial_\mu \pi \to \partial_\mu \pi + c_{\mu}$.
Since the DBI inflation is supposed to be
driven by the dynamics of the brane in the higher dimensional bulk,
it is interesting to study the effect of the Galileon-like terms 
in DBI inflation

In this paper, motivated by the DBI Galileons,
we studied  primordial fluctuations
generated during inflation described by the action
(\ref{Ginf}) which is obtained by
generalising the action of the DBI Galileons.
This generalisation is done in a similar way to the extension of 
the DBI inflation to the K-inflation.
In order to calculate the statistical quantities
of $\zeta$, the curvature perturbation on uniform density
hypersurfaces on large scales, at leading order in the
slow-varying approximations, we have adopted
the simple procedure \cite{Gruzinov:2004jx} to
first calculate the bispectrum of the fluctuations of inflaton $\phi$
in the flat gauge then relate it to that of $\zeta$
using the delta-N formalism (\ref{linear_rel_phi_zeta}).

For the linear perturbations, we have confirmed that,
owing to the Galileon-like term, the expression of
the sound speed $c_s$ for the scalar perturbations
is modified from the usual K-inflation model
(Eq.~(\ref{sound_speed})).
We also provided general expressions for the 
power spectrum $\mathcal{P}_\zeta$, spectral index $n_s$
and tensor to scalar ratio $r$ at leading order in the slow-varying 
approximations. In these expressions, $\tilde{\epsilon}$ defined by Eq.~(\ref{def_tilde_epsilon}), 
not $\epsilon \equiv -\dot{H}/H^2$, plays an important role (Eqs.~(\ref{power_spectrum_zeta}),
(\ref{spectral_index}) and (\ref{tensor_to_scalar})). Due to this, the consistency relation 
between the tensor to scalar ratio and the tensor spectrum index is broken if there 
exists the Galileon-like term.

We calculated the bispectrum at the leading order in the slow-varying variables. 
The Galileon-like term gives a new shape $\mathcal{A}_3$ in addition to the shapes
$\mathcal{A}_1$ and $\mathcal{A}_2$ which arise
in the usual K-inflation (Eqs.~(\ref{shape_func_a1}), (\ref{shape_func_a2})
and (\ref{shape_func_a3})).
For the new shape $\mathcal{A}_3$,
we checked the validity of using the
estimator for the equilateral-type Non-Gaussianity,  
$f_{\rm NL} ^{equil}$, based on the shape correlator
introduced by Ref.~\cite{Fergusson:2008ra} and
showed that the overlap is at about $99.99\%$ level,
which justifies the use of this estimator to measure the
amplitude of the bispectrum even in the presence of the
Galileon-like term. we obtained the general expression for the amplitude
of the bispectrum in Eq.~(\ref{fnl_gen_simplematching}).

For the concrete examples, we have considered two models:
one is the DBI Galileons described
by the action (\ref{relgal}).
The other is G-inflation model proposed by
Ref.~\cite{Kobayashi:2010cm}.
For the DBI Galileons, in the small sound speed limit,
$f_{\rm NL} ^{equil}$ is given by Eq.~(\ref{fnl_dbi}) and 
written in terms of the sound speed $c_s$ and $b_D$ that is related
to the amplitude of the Galileon-like term.
Since it scales as $f_{\rm NL} ^{equil} \propto 1/c_s^2$,
large primordial non-Gaussianities can be obtained
when $c_s$ is much smaller than $1$, similar to
the usual DBI inflation. It is worth mentioning that
for a given $c_s$,
the $b_D$-dependence of $f_{\rm NL} ^{equil}$
is weak and we obtained $-0.32 /c_s^2 < f_{\rm NL} ^{equil} < -0.16/c_s^2$. 
For G-inflation where an exact de Sitter solution is obtained without 
any potential terms, we found $f_{\rm NL} ^{equil}$ and the tensor to scalar ratio 
was related as $f_{\rm NL}^{equil} = 4.62 r^{-2/3}$. Although a small deviation from 
the de Sitter solution could give a correction to the tensor to scalar ratio, this 
relation gives a stringent constraint on the model by a detection or non-detection
of the equilateral type non-Gaussianity and the tensor to scalar ratio. 

In this paper, we considered the cubic order interaction in the Galileon theory and 
its relativistic generalisation. As is shown in \cite{deRham:2010eu}, it is possible to add 
two more higher order interactions which again arise from the probe brane action in a five-dimensional spacetime with the Gauss-Bonnet term. It is also possible to generalise these terms
in the same way as generalising Eq.~(\ref{relgal}) to Eq.~(\ref{Ginf}). 
It would be interesting to study phenomenology of
this generalisation. The single field model arises from a probe brane
action in the five-dimensional spacetime.
If the DBI Galileons have some connections to string theory, 
the DBI Galileons should be naturally a multi-field model
as in the DBI inflation where the position of the brane in each compact 
direction is described by a scalar field.
The multi-field Galileon model has been extensively studied recently
\cite{Hinterbichler:2010xn,Andrews:2010km,Deffayet:2010zh,
Padilla:2010de,Padilla:2010ir,Padilla:2010tj} and 
the relativistic extension of the model has been proposed 
\cite{Hinterbichler:2010xn}. We leave the study of multi-field 
DBI Galileons, the relativistic generalisation of the multi-field Galileon, 
for a  forthcoming paper.  

\begin{acknowledgments}
We would like to thank Shinji Tsujikawa and David Wands
for interesting discussions.
SM is supported by JSPS.
KK is supported by ERC, RCUK and STFC.

\end{acknowledgments}



\end{document}